\documentstyle[12pt,epsfig]{article}

\catcode`\@=11
\def\@citex[#1]#2{%
\if@filesw \immediate \write \@auxout {\string \citation {#2}}\fi
\@tempcntb\m@ne \let\@h@ld\relax \def\@citea{}%
\@cite{%
  \@for \@citeb:=#2\do {%
    \@ifundefined {b@\@citeb}%
      {\@h@ld\@citea\@tempcntb\m@ne{\bf ?}%
      \@warning {Citation `\@citeb ' on page \thepage \space undefined}}%
      {\@tempcnta\@tempcntb \advance\@tempcnta\@ne%
      \@tempcntb\number\csname b@\@citeb \endcsname \relax%
      \ifnum\@tempcnta=\@tempcntb 
	\ifx\@h@ld\relax%
	  \edef \@h@ld{\@citea\csname b@\@citeb\endcsname}%
	\else%
	  \edef\@h@ld{\ifmmode{-}\else--\fi\csname b@\@citeb\endcsname}%
	\fi%
      \else
	\@h@ld\@citea\csname b@\@citeb \endcsname%
	\let\@h@ld\relax%
      \fi}%
    \def\@citea{,\penalty\@highpenalty\,}%
  }\@h@ld
}{#1}}

\def\@citeb#1#2{{[#1]\if@tempswa , #2\fi}}
\def\@citeu#1#2{{$^{#1}$\if@tempswa , #2\fi }}
\def\@citep#1#2{{#1\if@tempswa , #2\fi}}

\def\bcites{         
	\catcode`\@=11
	\let\@cite=\@citeb
	\catcode`\@=12
}

\def\upcites{         
	\catcode`\@=11
	\let\@cite=\@citeu
	\catcode`\@=12
}

\def\plaincites{      
	\catcode`\@=11
	\let\@cite=\@citep
	\catcode`\@=12
}

\newcount\hour
\newcount\minute
\newtoks\amorpm
\hour=\time\divide\hour by 60
\minute=\time{\multiply\hour by 60 \global\advance\minute by-\hour}
\edef\standardtime{{\ifnum\hour<12 \global\amorpm={am}%
	\else\global\amorpm={pm}\advance\hour by-12 \fi
	\ifnum\hour=0 \hour=12 \fi
	\number\hour:\ifnum\minute<10 0\fi\number\minute\the\amorpm}}
\edef\militarytime{\number\hour:\ifnum\minute<10 0\fi\number\minute}

\def\draftlabel#1{{\@bsphack\if@filesw {\let\thepage\relax
   \xdef\@gtempa{\write\@auxout{\string
      \newlabel{#1}{{\@currentlabel}{\thepage}}}}}\@gtempa
   \if@nobreak \ifvmode\nobreak\fi\fi\fi\@esphack}
	\gdef\@eqnlabel{#1}}
\def\@eqnlabel{}
\def\@vacuum{}
\def\marginnote#1{}
\def\draftmarginnote#1{\marginpar{\raggedright\scriptsize\tt#1}}
\def\draft{
	\pagestyle{plain}
	\overfullrule=2pt
	\oddsidemargin -.5truein
	\def\@oddhead{\sl \phantom{\today\quad\militarytime} \hfil
	\smash{\Large\sl DRAFT} \hfil \today\quad\militarytime}
	\let\@evenhead\@oddhead
	\let\label=\draftlabel
	\let\marginnote=\draftmarginnote
	\def\ps@empty{\let\@mkboth\@gobbletwo
	\def\@oddfoot{\hfil \smash{\Large\sl DRAFT} \hfil}
	\let\@evenfoot\@oddhead}
	\def\@eqnnum{(\theequation)\rlap{\kern\marginparsep\tt\@eqnlabel}%
	\global\let\@eqnlabel\@vacuum}  }

\def\blackfonts{
	\font\blackboard=msbm10 scaled\magstep1
	\font\blackboards=msbm8
	\font\blackboardss=msbm6
}

\def\nblack{            
	\def\ZZ{{Z \n{10} Z}}
	\def\NN{{N \n{14} N}}
	\def\CC{{C \n{11} C}}
	\def\RR{{R \n{11} R}}
	\def\QQ{{Q \n{12} Q}}
	\def\PP{{P \n{11} P}}
}

\def\prep{         
	\catcode`\@=11
	\input art10.sty
	\catcode`\@=12
	
	\let\small\null
	\def\blackfonts{
		\font\blackboard=msbm10
		\font\blackboards=msbm7
		\font\blackboardss=msbm5
	}
	\let\sl\it
	\twocolumn
	\sloppy
	\voffset=-2.54truecm
	\hoffset=-2.54truecm
	\flushbottom
	\parindent 1em
	\leftmargini 2em
	\leftmarginv .5em
	\leftmarginvi .5em
	\marginparwidth 48pt
	\marginparsep 10pt
	\setlength{\columnsep}{2truecm}
	\setlength{\textwidth}{25.4truecm}
	\setlength{\textheight}{17truecm}
	\baselineskip=16pt
	\oddsidemargin .18truein
	\evensidemargin .17truein
}

\def\eqalign#1{\null\,\vcenter{\openup\jot\m@th
  \ialign{\strut\hfil$\displaystyle{##}$&$\displaystyle{{}##}$\hfil
      \crcr#1\crcr}}\,}
\def\eqalignno#1{\displ@y \tabskip\centering
  \halign to\displaywidth{\hfil$\@lign\displaystyle{##}$\tabskip\z@skip
    &$\@lign\displaystyle{{}##}$\hfil\tabskip\centering
    &\llap{$\@lign##$}\tabskip\z@skip\crcr
    #1\crcr}}

\def\section{\@startsection {section}{1}{\z@}{3.ex plus 1ex minus
 .2ex}{2.ex plus .2ex}{\large\bf}}
\def\subsection{\@startsection{subsection}{2}{\z@}{2.75ex plus 1ex minus
 .2ex}{1.5ex plus .2ex}{\bf}}

\def\appendix{{\newpage\section*{Appendix}}\let\appendix\section%
	{\setcounter{section}{0}
	\gdef\thesection{\Alph{section}}}\section}

\def\abstract{\if@twocolumn
\section*{Abstract}
\else 
\begin{center}
{\bf Abstract\vspace{-.5em}\vspace{0pt}}
\end{center}
\quotation
\fi}

\catcode`\@=12

\newcommand{\beq}{\begin{equation}}
\newcommand{\eeq}{\end{equation}}
\newcommand{\beqa}{\begin{eqnarray}}
\newcommand{\eeqa}{\end{eqnarray}}

\def\noj#1,#2,{{\bf #1} (19#2)\ }
\def\jou#1,#2,#3,{{\sl #1\/ }{\bf #2} (19#3)\ }
\def\ann#1,#2,{{\sl Ann.\ Physics\/ }{\bf #1} (19#2)\ }
\def\cmp#1,#2,{{\sl Comm.\ Math.\ Phys.\/ }{\bf #1} (19#2)\ }
\def\ma#1,#2,{{\sl Math.\ Ann.\/ }{\bf #1} (19#2)\ }
\def\ng#1,#2,{{\sl Nagoya.\ Math.\ J.\/ }{\bf #1} (19#2)\ }
\def\jd#1,#2,{{\sl J.\ Diff.\ Geom.\/ }{\bf #1} (19#2)\ }
\def\invm#1,#2,{{\sl Invent.\ Math.\/ }{\bf #1} (19#2)\ }
\def\cq#1,#2,{{\sl Class.\ Quantum Grav.\/ }{\bf #1} (19#2)\ }
\def\cqg#1,#2,{{\sl Class.\ Quantum Grav.\/ }{\bf #1} (19#2)\ }
\def\ijmp#1,#2,{{\sl Int.\ J.\ Mod.\ Phys.\/ }{\bf A#1} (19#2)\ }
\def\jmphy#1,#2,{{\sl J.\ Geom.\ Phys.\/ }{\bf #1} (19#2)\ }
\def\jams#1,#2,{{\sl J.\ Amer.\ Math.\ Soc.\/ }{\bf #1} (19#2)\ }
\def\grg#1,#2,{{\sl Gen.\ Rel.\ Grav.\/ }{\bf #1} (19#2)\ }
\def\mpl#1,#2,{{\sl Mod.\ Phys.\ Lett.\/ }{\bf A#1} (19#2)\ }
\def\nc#1,#2,{{\sl Nuovo Cim.\/ }{\bf #1} (19#2)\ }
\def\np#1,#2,{{\sl Nucl.\ Phys.\/ }{\bf B#1} (19#2)\ }
\def\pl#1,#2,{{\sl Phys.\ Lett.\/ }{\bf #1B} (19#2)\ }
\def\pla#1,#2,{{\sl Phys.\ Lett.\/ }{\bf #1A} (19#2)\ }
\def\pr#1,#2,{{\sl Phys.\ Rev.\/ }{\bf #1} (19#2)\ }
\def\prd#1,#2,{{\sl Phys.\ Rev.\/ }{\bf D#1} (19#2)\ }
\def\prl#1,#2,{{\sl Phys.\ Rev.\ Lett.\/ }{\bf #1} (19#2)\ }
\def\prp#1,#2,{{\sl Phys.\ Rept.\/ }{\bf #1C} (19#2)\ }
\def\ptp#1,#2,{{\sl Prog.\ Theor.\ Phys.\/ }{\bf #1} (19#2)\ }
\def\ptpsup#1,#2,{{\sl Prog.\ Theor.\ Phys.\/ Suppl.\/ }{\bf #1} (19#2)\ }
\def\rmp#1,#2,{{\sl Rev.\ Mod.\ Phys.\/ }{\bf #1} (19#2)\ }
\def\yadfiz#1,#2,#3[#4,#5]{{\sl Yad.\ Fiz.\/ }{\bf #1} (19#2) #3%
\ [{\sl Sov.\ J.\ Nucl.\ Phys.\/ }{\bf #4} (19#2) #5]}
\def\zh#1,#2,#3[#4,#5]{{\sl Zh.\ Exp.\ Theor.\ Fiz.\/ }{\bf #1} (19#2) #3%
\ [{\sl Sov.\ Phys.\ JETP\/ }{\bf #4} (19#2) #5]}

\def\beq{\begin{equation}}
\def\eeq{\end{equation}}
\def\beqar{\begin{eqnarray}}
\def\eeqar{\end{eqnarray}}

\def\hE{\widehat{E}}

\newcommand{\be}{\begin{equation}}
\newcommand{\ee}{\end{equation}}
\newcommand{\bea}{\begin{eqnarray}}
\newcommand{\eea}{\end{eqnarray}}

\def\nfrac#1#2{{\displaystyle{\vphantom1\smash{\lower.5ex\hbox{\small$#1$}}%
	\over\vphantom1\smash{\raise.25ex\hbox{\small$#2$}}}}}

\def\p#1{\mskip#1mu}
\def\n#1{\mskip-#1mu}
\def\stop{\p6.}
\def\comma{\p6,}

\def\to{\rightarrow}

\def\lae{\mathrel{\mathop{\smash{\lower .5 ex \hbox{$\stackrel<\sim$}}}}}
\def\lae{\mathrel{\mathop{\smash{\lower .5 ex \hbox{$\stackrel>\sim$}}}}}

\def\l:{\mathopen{:}\,}
\def\r:{\,\mathclose{:}}

\catcode`\@=11
\def\theequation{\arabic{equation}}
\catcode`\@=12

\nblack
\bcites

\nblack

\catcode`\@=11
\def\theequation{\thesection.\arabic{equation}}
\@addtoreset{equation}{section}
\@addtoreset{footnote}{section}
\@addtoreset{footnote}{subsection}
\catcode`\@=12

\newcommand{\beqn}{\begin{equation}}
\newcommand{\eeqn}{\end{equation}}
\newcommand{\beqnarray}{\begin{eqnarray}}
\newcommand{\eeqnarray}{\end{eqnarray}}
\newcommand {\bear} [1] {\begin {array} {#1}}
\newcommand {\ear} {\end {array}}

\newcommand {\beqarn} {\begin{eqnarray*}}
\newcommand {\eeqarn} {\end{eqnarray*}}

\begin{document}
\begin{titlepage}

\begin{center}
\today
\hfill LBNL-40031, UCB-PTH-97/10 \\
\hfill                  hep-th/9702173

\vskip 1.5 cm
{\large \bf  F-Theory, T-Duality on  $K3$ Surfaces and 
$N=2$ Supersymmetric Gauge Theories
in Four Dimensions}
\vskip 1 cm 
{Kentaro Hori and Yaron Oz }\\
\vskip 0.5cm
{\sl Department of Physics,
University of California at Berkeley\\
366 Le\thinspace Conte Hall, Berkeley, CA 94720-7300, U.S.A.\\
and\\
Theoretical Physics Group, Mail Stop 50A--5101\\
Ernest Orlando Lawrence Berkeley National Laboratory, 
Berkeley, CA 94720, U.S.A.\\}

\end{center}

\vskip 0.5 cm
\begin{abstract}
We construct T-duality on $K3$ surfaces. The T-duality
exchanges a 4-brane R-R charge and a 0-brane R-R charge.
We study the action of the T-duality on the moduli space
of 0-branes located at points of $K3$ and 4-branes wrapping it.
We apply the construction to F-theory compactified on a Calabi-Yau 4-fold and
study the  duality of $N=2$ $SU(N_c)$ gauge theories
in four dimensions.
We discuss the generalization to the $N=1$ duality scenario.

\end{abstract}

\end{titlepage}

\section{Introduction}

The string interpretation of the duality between four dimensional $N=1$ supersymmetric 
gauge theories  has been studied recently \cite{b,vz,david,ov}.
It has been suggested in \cite{b} that 
the duality between four dimensional $N=1$ supersymmetric 
gauge theories \cite{s,k} may be understood as a consequence
of T-duality in string theory.
The crucial point for understanding the $N=1$ duality
in this framework is the meaning
of T-duality of a K\"ahler surface which is not a torus
and is embedded in a Calabi-Yau space.
Our aim in this paper is to try to gain an understanding of the
required generalization
of the notion of T-duality and its implications.

The framework for studing the duality phenomena
will be the same as suggested in \cite{b}.   
Consider a compactification of $F$-theory on a Calabi-Yau 4-fold
elliptically fibered
over a 3-fold base $B$.
 This leads to
an $N=1$ theory in four dimensions.
Let $S$ be a complex surface in $B$
along which the elliptic fibration acquires singularity of
the $A_{N_c-1}$ type.
We consider a 7-brane with 
worldvolume ${\bf R}^4 \times S$ on which we have
 an $SU(N_c)$ gauge symmetry.
In addition there are $h^{1,0}(S) + h^{2,0}(S)$ 
chiral multiplets in the adjoint representation.
We will also add $N_f$ 3-branes with world volume
${\bf R}^4$ which are located at points of the 
 surface $S$. The open strings stretching between
the 3-branes and the 7-brane give
 $N_f$  hypermultiplets in the fundamental
representation of the gauge group.

The Higgs branch of the supersymmetric gauge theory on ${\bf R}^4$
is constructed as the moduli space 
of 0-branes and 4-branes on $S$.
T-duality maps this moduli space to another D-brane moduli space
which describes the Higgs
branch of the dual theory. 
In section 2 we will begin by defining the D-brane moduli space
as a space of vector bundles on $S$.
In particular we will see that we are forced to generalize
the notion of a vector bundle
to that of a sheaf, as suggested in \cite{HM}.
We will discuss the  modification
for the study of the D-brane moduli space
when $S$ is embedded in a curved space.
In section 3 we will construct a generalization of T-duality for
$K3$ surfaces,
which maps a 0-brane charge to a 4-brane charge and vice versa.
We will study its properties,
 check its consistency with the duality between
the heterotic string on $T^4$ and
type IIA string theory on $K3$,
and  compare it to the mirror transform of $K3$.
We will then study the implications to the $N=2$ duality.
Finally, we will discuss  the case when $S$ is a rational surface,
which is the relevant
surface for the study of $N=1$ duality.

\section{D-Brane Moduli Space}
Let us consider type II string theory compactified on
a manifold $X$ of real dimension $2d$.
We are interested in the moduli space of D-branes wrapping
supersymmetric cycles in $X$.
BPS states are associated with the cohomology classes
of the D-brane moduli space.
Consider a configuration of $2d$-branes wrapped on $X$.
It carries charges for various RR fields which,
as shown in \cite{I,HM}, takes the following form
\beq
Q= v(E) = {\rm ch}(E)\sqrt{\widehat{\rm A}(X)}
\stop
\label{ve}
\eeq
${\rm ch}(E)$ is the Chern character of the vector bundle $E$ 
\beq
{\rm ch}(E) = {\rm Tr} \exp\left[\frac{1}{2\pi}(F-B)\right]
\comma
\label{che}
\eeq
where $F$ is the field strength of the gauge field on the brane and $B$
is the bulk
NS-NS 2-form.
It has an expansion in terms of the Chern classes 
\beq
{\rm ch}(E) 
= {\rm rank}(E) + c_1(E) + \frac{1}{2}(c_1^2(E) - 2c_2(E)) + ...
\stop
\label{ee}
\eeq
$\widehat{\rm A}(X)$
is the A-roof genus and it has an expansion in terms of the
Pontrjagin classes
\beq
\widehat{\rm A}(X) = 1 - \frac{p_1(X)}{24} + ...
\stop
\label{eA}
\eeq

$v(E)$ is what is known as the Mukai vector
of the vector bundle $E$ on $X$
\footnote{In \cite{Mukai1}, the Mukai vector is defined as
$v(E) = {\rm ch}(E)\sqrt{{\rm Td}(X)}$. 
This coincides with (\ref{ve}) when $X$ is a Calabi-Yau space.}.
In the following we will be interested in the case where $X$ is a complex
surface.
In this case $v(E) \in H^0(X,{\bf Z})
\oplus H^2(X,{\bf Z}) \oplus H^4(X,{\bf Z})$, and
expanding (\ref{che}) using (\ref{ee}) and (\ref{eA}) 
 we have
\beq
v(E) = \Bigl({\rm rank}(E), c_1(E),
\frac{1}{2} c_1^2(E) - c_2(E) - \frac{p_1(X)}{48} {\rm rank}(E)\Bigr)
\stop
\eeq

Consider now one 4-brane wrapped on $X$.
It corresponds  to  a flat $U(1)$ bundle on $X$.
However, if $p_1(X) \neq 0$ the 4-brane
induces a 0-brane charge via the term $\frac{1}{48}\int_X p_1(X) A_1$ 
in its effective action, where $A_1$ is the RR 1-form.
Indeed, the  Mukai vector corresponding to a 4-brane is
$v(E) = (1,0,-\frac{p_1}{48})$.
In this paper, we take the convention that the charge vector of the
0-brane is $(0,0,-1)$.
For instance, after integrating $p_1(X)$ over the surface $X$
the Mukai vector for a 4-brane wrapping $T^4$ is $v(E) = (1,0,0)$, 
while the Mukai vector for a 4-brane wrapping $K3$ is $v(E) = (1,0,1)$
and induces the 0-brane charge $-1$.

The D-brane moduli space can
be viewed as the moduli space of vector bundles $E$
on $X$. To be more precise,
we need to consider not only vector bundles
but also coherent sheaves\footnote{
It has been advocated in \cite{HM} that
the appropriate objects are coherent simple semistable sheaves.}.
A {\it coherent sheaf} on $X$ is represented as a cokernel of
a map of vector bundles
on $X$. A notable difference between
coherent sheaves and vector bundles is that while the dimension
of the fiber of a vector bundle
is constant as we move along the base $X$,
the dimension of the fiber of a coherent sheaf is allowed to jump.
For illustration, consider a configuration with
one 4-brane wrapped on a $K3$ surface $X$ and
$n$ 0-branes at points in $X$.
It has the charge vector $(1,0,1-n)$.
There is no vector bundle whose Mukai vector is $v(E) = (1,0,1-n)$,
namely no line bundle can have non-zero second Chern number $n$.
But there is indeed such a sheaf.
It is a sheaf of holomorphic functions on $X$
vanishing at $n$ points.
(This is an element of the so called {\it Hilbert scheme of
$n$-points in $X$}.)
This simple example indicates that
the use of this generalized notion of a vector bundle
enables us to describe the D-brane moduli spaces
of various charges on the same footing,
including those whose charge vector is
not realized as the Mukai vector of a vector bundle.
As to terminology,  we will still use the notion of  vector bundles,
although it should 
be clear from the above that in some of the cases the objects are really 
coherent sheaves. 

A 0-brane looks like a zero size instanton
on a 4-brane wrapping $S$ \cite{W,D,V}. 
While coherent sheaves are objects of
algebraic geometry, instantons are objects of differential geometry.
However, the intuitive relation between small instantons
and coherent sheaves is correct \footnote{Small instantons are needed
for the (Uhlenbeck)
compactification
of the instanton moduli space, while the coherent sheaves are needed
for the (Gieseker)
compactification of the moduli space of stable vector bundles, and
on algebraic complex surfaces the two compactifications
are related \cite{JLi}.}.

Let us consider
D branes (partially) wrapped on
cycles in a manifold $X$ which is embedded in a curved manifold.
In particular, $X=S$ in the base $B$ of an elliptic Calabi-Yau 4-fold
defining an F-theory vacuum.
Then the formula (\ref{ve}) for the RR charge vector
will be in general modified.
In such a case 
the scalar and fermionic fields on the worldvolume of
the brane are in general twisted \cite{BSV}.
If $X$ was embedded in a manifold for type II compactification,
the scalars would be sections of the normal bundle while the fermions
would be sections of the spin bundle
tensored by the square root of the normal bundle.
Since the normal bundle to the worldvolume
of the brane is in general non-trivial the scalars
and the fermions are twisted.

In the framework that we want to study, in which
$X=S$ embedded in the base $B$ of F-theory compactification,
we do not know in detail how to twist the fields. 
Nevertheless the
twist can be uniquely determined \cite{KV}.
On a flat 7-brane, we would have the $N=1$ supersymmetry
in eight dimensions.
Our requirement is to have $N=1$
supersymmetry on the uncompactified direction
${\bf R}^4$ of the 7-brane wrapped on $S\times {\bf R}^4$.
On a K\"ahler manifold with spin structure, spinors are $(0,p)$ forms 
with values in
the square root of the canonical line bundle $K^{\frac{1}{2}}$.
This implies that we twist the fermions by $K^{-\frac{1}{2}}$
and therefore  they transform as $(0,p)$ forms.
For $X$ being $T^4$ or $K3$
the canonical class is trivial and therefore (\ref{ve}) is not modified.
This is not the case
for the rational surfaces which are of interest to us for the
case of $N=1$ duality.
For example, for the Hirzebruch surface $S$ with $p_1(S)=0$,
the formula (\ref{ve}) without modification would show that the 4-brane
does not induce 0-brane charge and that
T-duality proposed in \cite{b} does not lead to $N=1$ duality.


\section{$N=2$ Duality}

\subsection{Fourier-Mukai Transform for $K3$ }

Our aim is to generalize the concept of T-duality to surfaces other
than $T^4$.
In this section we will construct a generalization of T-duality for
$K3$ surfaces.
The generalization will be a natural extension
of the Nahm transform  \cite{BB,S}
which is
a way of viewing T-duality on $T^4$ in the differential
geometric language, and is known as the Fourier-Mukai 
transform \cite{Mukai} in the algebraic geometry framework.

Let us first discuss  T-duality on $T^4$ and 
the action of T-duality on the moduli space of D-branes on
$T^4$. In particular we are interested in the action of T-duality
on 0-branes located at points on 
the $T^4$ and 4-branes wrapping it.
In the language of the previous section the torus is the 
moduli space of a 0-brane
on $T^4$ with charge vector $(0,0,-1)$.
The dual torus $\widehat{T}^4$ is the moduli space
of flat $U(1)$ bundles on 
$T^4$ or line bundles on $T^4$ with
Mukai vector $v=(1,0,0)$. 
In other words, the 
dual torus $\widehat{T}^4$ is the moduli space of a 4-brane wrapping $T^4$.
Given a vector bundle $E$  on $T^4$ which describes a configuration of
D-branes on $T^4$, the dual
 bundle $\widehat{E}$ on $\widehat{T}^4$ 
 is constructed as the (negative) index bundle $-{\rm Ind}D$
of a family of Dirac operators $D_{\hat{t}}$ associated with
 the twisted vector bundles
$E_{\hat{t}} =  E\otimes L_{\hat{t}}$. $L_{\hat{t}}$
are line bundles on $T^4$ with Mukai vector $(1,0,0)$
 parametrized by the dual torus, $\hat{t}\in \widehat{T}^4$.
One can compute the Mukai vector of $\widehat{E}$ by using
the family index theorem 
 \beq
 {\rm ch}({\rm Ind}D)
 = \int_{T^4} {\rm ch}(E\otimes {\cal Q})\widehat{\rm A}(T^4)
 \comma
 \label{index}
 \eeq
where ${\cal Q}$ is the so called Poincar\'e bundle over
$T^4\times \widehat{T}^4$
such that its restriction on $T^4\times \{\hat{t}\}$ is $L_{\hat{t}}$.
As computed explicitly
in \cite{BB,DK}, for $c_1(E)=0$ we have
 \beq
 {\rm rank}(\widehat{E}) = c_2(E),~~~~~~c_2(\widehat{E}) = {\rm rank}(E)
 \stop
 \eeq
 This is what we expect from T-duality under which 0-branes
and 4-branes are exchanged.
 
 In order to generalize the above construction
of T-duality to $K3$ we first have to define 
 the dual $K3$. There are many ways to define the dual $K3$ \cite{Mukai1}
 but only one corresponds to
 the required T-duality on all four coordinates.\footnote{
  A Fourier-Mukai transform for reflexive
  $K3$ surfaces has been derived in a rigorous way in \cite{it}.
  However, the case studied
in that paper does not correspond to the required T-duality.}
 Later we will 
   construct  for comparison
 the dual $K3$ that is obtained by a  mirror transform.

We can view $K3$ as the moduli space of a 0-brane on $K3$
with RR charge vector $(0,0,-1)$.
 Naively we may think that the dual $K3$ is the moduli space
of a 4-brane wrapping
  $K3$.
 This cannot be correct on dimensional ground.
 The complex dimension of the moduli space of vector bundles on $K3$ with 
 Mukai
 vector $v=(r,l,s)$ is $l^2 -2rs +2$ \cite{Mukai1}.
 As we saw in the previous section,
 the Mukai vector of a 4-brane wrapping
$K3$ is $v= (1,0,1)$ and  the dimension
 of the moduli space of a 4-brane wrapping $K3$ is zero,
thus it cannot be a dual to $K3$.

 Indeed, in analogy with the torus
 case,
  the correct dual should be the moduli space of sheaves
    with Mukai vector $v=(1,0,0)$. Such a Mukai vector
      corresponds to 
 one 0-brane and one 4-brane.
 This means that T-duality on $K3$ does not map a 0-brane to a 4-brane,
 but rather
 a 0-brane to a 4-brane plus a 0-brane.
 In other words T-duality on $K3$ does not map a physical
 0-brane to a physical 4-brane but rather a 0-brane charge
to a 4-brane charge,
 and vice versa.
A sheaf with Mukai vector $(1,0,0)$
has rank one, $c_1=0$ and $c_2=1$. It cannot be
a vector (line) bundle.
Rather, as remarked previously,
it is a sheaf of holomorphic functions vanishing
at a point. By assigning such a point to each sheaf, we obtain
a bijection of the moduli space of sheaves with Mukai vector
$(1,0,0)$ to the original $K3$.
This is the Hilbert scheme of one point on $K3$.

 Given a vector bundle $E$  on a $K3$ surface $X$ which describes
 a configuration of D-branes on $X$, we wish to
 construct the dual
 bundle $\widehat{E}$ as the (negative) index bundle of a Dirac operator
associated with $E_{\hat{x}} =  E\otimes L_{\hat{x}}$ where $L_{\hat{x}}$
are sheaves on $X$
with Mukai vector $(1,0,0)$
 parametrized by $\hat{x}\in \widehat{X}$.
 However, as $L_{\hat{x}}$ is not locally free (i.e. not a vector bundle),
 it is not obvious how to define the Dirac operator. 
Now we recall that on a $K3$ surface, the positive and negative
spin bundles are
$S_+=\Omega^{0,0}\oplus\Omega^{0,2}$ and $S_-=\Omega^{0,1}$
respectively,
where $\Omega^{0,p}$ is the bundle of anti-holomorphic $p$-forms,
and the Dirac operator is essentially the $\overline{\partial}$ operator.
Thus, the index bundle of the Dirac operator can be considered
as the bundle
of Dolbeault cohomology groups $H^{0,0}-H^{0,1}+H^{0,2}$.
With a twisted coefficient ${\cal E}$, this is the same as the bundle of
cohomology groups $H^0(X,{\cal E})-H^1(X,{\cal E})+H^2(X,{\cal E})$,
which can be extended to the case where ${\cal E}$ is not locally free.
Applying to the case ${\cal E}=E_{\hat{x}}$, we can define the dual bundle
$\widehat{E}$ as such an index bundle with its sign inverted.

Applying the Grothendieck-Riemann-Roch theorem,
which is an analog of
the family index theorem,
we can compute the Chern character of $\widehat{E}$:
\beq
{\rm ch}(-\widehat{E})=\int_X{\rm ch}(E\otimes {\cal Q}){\rm Td}(X).
\label{GRR}
\eeq
The Poincar\'e bundle ${\cal Q}$ is a bundle on $X\times \widehat{X}$
such that the restriction to $X\times\{\hat{x}\}$ is $L_{\hat{x}}$.
It is a sheaf of holomorphic functions on $X\times\widehat{X}$
vanishing on the diagonal $\Delta\cong X$ (recall that $\widehat{X}$ is
canonically isomorphic to $X$).
Since the restriction
of ${\cal Q}$ to $X\times \{\hat{x}\}$ is $L_{\hat{x}}$
whose Chern character is
$1-w_X$ where $w_X$ is the 4-form of $X$ with volume one,
${\rm ch}({\cal Q})$ must have the term $1-w_X$
(pulled back to $X\times\widehat{X}$). Similarly, it must have
$1-w_{\widehat{X}}$ and thus,
it must contain the term $1-w_X-w_{\widehat{X}}$.
For the purpose of our calculation,
we want to know the coefficient of 
the term $w_Xw_{\widehat{X}}$ in ${\rm ch}({\cal Q})$.
Note that we have an exact sequence of sheaves
\beq
0\longrightarrow
{\cal Q}\longrightarrow
{\cal O}_{X\times \widehat{X}}\longrightarrow
{\cal O}_{\Delta}\longrightarrow 0
\comma
\label{Qseq}
\eeq
where ${\cal O}_{X\times \widehat{X}}$ is the sheaf of holomorphic
functions on $X\times \widehat{X}$, and ${\cal O}_{\Delta}$
is the sheaf of holomorphic functions supported on $\Delta$. 
From this we obtain 
$\chi(X\times \widehat{X},{\cal O})
=\chi(X,{\cal O})+\chi(X\times\widehat{X},{\cal Q})$.
Since $h^{0,0}=h^{0,2}=1$ and $h^{0,1}=0$ for
$X\cong \widehat{X}$, we have $\chi(X,{\cal O})=2$, and thus, we see that
$\chi(X\times\widehat{X},{\cal Q})=2$. Applying the Riemann-Roch formula
$\chi(X\times\widehat{X},{\cal Q})=
\int_{X\times\widehat{X}} {\rm ch}({\cal Q}){\rm Td}(X\times\widehat{X})$,
and using
${\rm Td}(X)=1+2w_X$ and ${\rm Td}(\widehat{X})=1+2w_{\widehat{X}}$
together with the property
${\rm Td}(X\times\widehat{X})={\rm Td}(X){\rm Td}(\widehat{X})$,
we see that
\beq
{\rm ch}({\cal Q})=1-w_X-w_{\widehat{X}}+2w_Xw_{\widehat{X}}
\eeq
up to a possible term in $H^2(X)\wedge H^2(\widehat{X})$
which does not contribute to the index for $c_1(E)=0$.
Then, the formula (\ref{GRR}) yields
$-{\rm ch}(\widehat{E})
={\rm ch}_2(E)+{\rm rank}(E)-w_{\widehat{X}}{\rm ch}_2(E)$
in the case $c_1(E)=0$.
Namely, we have seen that  

\beq
 {\rm rank}(\widehat{E}) 
 = c_2(E) - {\rm rank}(E),~~~~~~c_2(\widehat{E}) = c_2(E) 
 \stop
 \label{M}
 \eeq

Equation (\ref{M}) describes the action of T-duality on the moduli space of 4-branes 
wrapping the $K3$ surface and 0-branes located at points on it.

It is instructive, for a comparison with T-duality,
to define mirror symmetry of $K3$ surfaces
in the above language.
Following  \cite{Ko,SYZ,Morrison,bb}, we  define the mirror of $K3$  
as the moduli space of 2-branes wrapping supersymmetric 2-cycles with a
topology 
of $T^2$ in $K3$ (holomorphically embedded elliptic curves).
The Mukai vector for such a brane is $v=(0,u,0)$  where  $u^2=0$
is the self-intersection number of $T^2$ in $K3$.
Since $K3$ can be viewed as the moduli space of a 0-brane on it,
with Mukai vector $v=(0,0,-1)$
we see that
mirror symmetry transforms $v=(0,0,-1)\rightarrow v=(0,u,0), u^2=0$.
Given a bundle
$E$ on $K3$ describing a configuration of D-branes, the dual bundle
$\hE$ can be constructed  as before as the index bundle of $E_{\hat{x}}$
and has a ${\rm rank}(\hE) = -c_1(E) u$.
For instance the 2-brane with Mukai vector $(0,u,0)$, $u^2=-2$,
wrapping a rational curve (supersymmetric 2-cycle $S^2$)
which intersect with the $T^2$ transversally
is mapped to a 4-brane with Mukai vector $(1,0,0)$.

The duality between
the heterotic string on $T^4$ and the type IIA string on $K3$
is a useful way to gain some further understanding
 of the meaning of T-duality on $K3$.
 We will now show that a particular T-duality on $T^4$ at the heterotic
 side corresponds to the above T-duality on $K3$.
The integer homology lattice of $K3$
can be decomposed as $\Gamma_{3,19}\oplus \Gamma_{1,1}$
where  $\Gamma_{3,19}$ corresponds to $H_2(K3,{\bf Z})$ 
and $\Gamma_{1,1}$  to $H_0(K3,{\bf Z})\oplus H_4(K3,{\bf Z})$.
We can decompose that Narain lattice $\Gamma_{4,20}$
in the heterotic side in a similar way as
$\Gamma_{3,19}\oplus \Gamma_{1,1}$, and let $(p_R,p_L)$ denote
the momenta in the $ \Gamma_{1,1}$
part.

T-duality on the torus maps $p_R \pm p_L \rightarrow  p_R\mp p_L$.
We argued that T-duality
on $K3$ exchanges 0-brane charge and 4-brane charge. 
It is natural to ask whether T-duality on $T^4$  and T-duality on $K3$
are consistent with the heterotic-type IIA duality.
This will be the case if  $\frac{1}{\sqrt{2}}(p_R + p_L)$ 
corresponds to 0-brane charge and
$\frac{1}{\sqrt{2}}(p_R - p_L)$
corresponds to 4-brane charge, or vice versa.
It is easy to see that this is correct.
The product  $\frac{1}{2}(p_R+ p_L)(  p_R - p_L)$ is the length
of a vector $(p_R,p_L)$ in
$\Gamma_{1,1}$. This is mapped 
by the heterotic-type IIA duality to the intersection number
of 0-branes and 4-branes
on $K3$, or more accurately
taking into account the induced 0-brane charge from a 4-brane on $K3$,
to   the product of 0-brane charge and 4-brane charge \cite{vafa}.
Thus we see that the dual $K3$ that we constructed is natural
from the viewpoint
of string duality.

Note that since the construction of the dual $K3$
is not affecting the $\Gamma_{3,19}$ lattice
of $K3$, it is natural to expect that
the T-duality on $K3$ preserves its complex structure.
We have already observed this since the Hilbert scheme of one point
on $X$ is the same as
$X$ itself $\widehat{X}\cong X$.
Note that when constructing
the mirror to $K3$ we also affect the $\Gamma_{3,19}$ part
of the lattice and therefore
change the complex structure, in accord with the mirror transform. 

Let us now discuss what happens to the volume of $K3$ after T-duality.
We expect that the volume
of the dual $K3$ will be proportional
to the inverse of the original $K3$.
In order to show that consider the decomposition of a vector
$B' \in {\bf R}^{4,20}$ as
\cite{asp}  
\beq
B' = \alpha w + w^* + B
\comma
\label{w}
\eeq
where  $B\in {\bf R}^{3,19}$ is the NS-NS two form, and
$~w, w^* \in \Gamma_{1,1}$ satisfy
$w\cdot w = w^* \cdot w^* = 0, w^* \cdot w =1$.
It is argued in \cite{asp} that $\alpha$ is the volume of the $K3$ surface.
T-duality for $K3$ as constructed above exchanges  $w^* \leftrightarrow w$
and indeed, as seen from (\ref{w}), it inverts the volume
of the $K3$ surface
$\alpha\to1/\alpha$, as expected. 

In closing this section let us comment  how we can see from the orbifold
viewpoint that the T-duality maps 4-brane charge to 0-brane charge and not
physical 4-branes
to physical 0-branes.
On the surface $X$ the coupling to the R-R 1-form $A_1$ and 5-form $A_5$
has the structure 
\beq
(-c_2(E)  - \frac{p_1(X)}{48})A_1 + {\rm rank} E \int_X A_5
\comma
\label{ch}
\eeq
where the term multiplying $A_1$ is the 0-brane charge while the term
multiplying $\int_X A_5$ is the 4-brane charge.
When $X$ is an orbifold we can still use flat coordinates. In particular,
 the R-R forms are constructed
using the zero modes $\frac{1}{2}(\psi_0^{\mu} \pm \tilde{\psi_0}^{\mu})$.
T-duality maps $\frac{1}{2}(\psi_0^{\mu} \pm \tilde{\psi_0}^{\mu})
\rightarrow \frac{1}{2}(\psi_0^{\mu} \mp \tilde{\psi_0}^{\mu})$.
This exchanges the R-R fields $A_1$ with $A_5$, and since the (\ref{ch})
has to be preserved (if T-duality is a symmetry)
the 4-brane and 0-brane charges must be exchanged.

\subsection{$N=2$ Duality}

When $S=K3$,
since $h_{2,0}(K3)=1$ we get an $N=2$ supersymmetry in
the uncompactified direction ${\bf R}^4$ of the worldvolume
of the 7-brane wrapping $S\times {\bf R}^4$. 
We can approximate the F-theory configuration near the 7-brane 
by a perturbative 
type IIB string theory compactified on $K3$
with parallel $N_c$ 7-branes wrapped on
$K3\times {\bf R}^4$.
Indeed, such a configuration
 yields
$N=2$ supersymmetry on the uncompactified direction ${\bf R}^4$
of the worldvolume.  
The gauge group is $SU(N_c)$ and the matter content is $N_f$
hypermultiplets in the 
fundamental representation.

In the model that we consider the D-brane moduli space  
describes vector bundles $E$
with  ${\rm rank}(E) = N_c, c_1(E) =0, c_2(E) = N_f$.
In principle, there is another gauge group $U(N_f)$ corresponding to the 
$N_f$ 3-branes.
However, we are looking at worldvolume dynamics of the 7-brane.
Thus, the $U(N_f)$ group appears in this framework  as a global symmetry.

In the following discussion,
neglecting the uncompactified direction ${\bf R}^4$ for a while,
we will use the words 4-branes and 0-branes instead of
7-branes and 3-branes respectively.
The Mukai vector describing 
$N_c$ 4-branes wrapping $K3$ and $N_f$ 0-branes located at points on $K3$
is
\beq
v(E) = (N_c, 0, N_c-N_f)
\stop
\label{mu}
\eeq

The moduli space of $N_c$ 4-branes wrapping $S$ and $N_f$ 0-branes
located at points on $S$
is the moduli space of vector bundles on $K3$ with
Mukai vector (\ref{mu}).
The complex dimension of this space is 
\beq
\dim \,M_{v=(N_c,0,N_c-N_f)}(K3)= 2N_cN_f - 2(N_c^2 -1)
\stop
\label{dim}
\eeq
The description of 0-branes on the 4-branes
as instantons suggests that the moduli space of $N_f$ 0-brane on
$N_c$ 4-branes wrapping $K3$ 
$M_{v=(N_c,0,N_c-N_f)}(K3)$
is  closely related
to the moduli space of $SU(N_c)$  $N_f$-instantons on $K3$.

The link between the D-branes and
the supersymmetric gauge theory in ${\bf R}^4$ is the
identification of the D-brane moduli space
and the Higgs branch of the gauge theory.
This presumably requires some limit such as large volume of the surface.
The Higgs branch of $N=2$ $SU(N_c)$ gauge theory
with $N_f$ hypermultiplets in the fundamental
representation 
contains two kinds of branches:
The Baryonic branch and the non-Baryonic branch \cite{aps}. 
Only in the Baryonic branch the gauge group is completely
Higgsed and one has a pure Higgs
branch. The non Baryonic branch extends to a mixed branch.
The Baryonic and non-Baryonic branches intersect classically,
and are separated  due to instanton
correction in the quantum theory.
On dimensional ground,
we expect that the D-brane moduli space  describes the Baryonic branch.

Using the results of the previous section (\ref{M}),
T-duality on $K3$ maps the Mukai vector
(\ref{mu}) to   
\beq
v(\hE) = (N_f - N_c, 0, -N_c)
\stop
\label{mu1}
\eeq

The moduli space of D-branes on $K3$ and the moduli space of D-branes
on the dual $K3$ are
isomorphic. Thus, the T-duality
 suggests that the Baryonic branch of 
  $N=2$ $SU(N_c)$ gauge theory with $N_f$ hypermultiplets
in the fundamental representation
  is identical to the Baryonic
  branch of 
  $N=2$ $SU(N_f- N_c)$ gauge theory with $N_f$ hypermultiplets
in the fundamental representation.

The Higgs branch of $N=2$ supersymmetric QCD was studied in \cite{Anto} where
 it was claimed that the part of
the moduli space corresponding to complete Higgsing
(open dense subset of
the Baryonic branch) of $N=2$ $SU(N_c)$ SQCD with $N_f$ flavors
is given by the cotangent bundle of
the total space
of the determinant line bundle of the Grassmannian
$Gr(N_c,N_f)$ with its zero section deleted.
This claim is correct up to a subtle point, which we will clarify in the following.

Let us denote by $Q, \widetilde{Q}$ the pair of $N=1$ chiral superfields that 
constitute
a hypermultiplet of $N=2$ supersymmetry. Here we consider $\widetilde{Q}$ 
as a map from ${\bf C}^{N_c}$ to ${\bf C}^{N_f}$ and $Q$
as a map from ${\bf C}^{N_f}$ to ${\bf C}^{N_c}$.
The Higgs branch is constructed as the set of $SL(N_c,{\bf C})$ orbits
of solutions of the F-flatness equation
\beq
Q\widetilde{Q} \propto {\bf 1}_{N_c}
\stop
\label{hk}
\eeq
When ${\rm rank}\widetilde{Q}=N_c$,
$\widetilde{Q}$ defines a non-zero point in the determinant line
bundle of $Gr(N_c,N_f)$.
Then, $Q$ defines a linear form
$\delta \widetilde{Q}\mapsto {\rm Tr}(Q\delta\widetilde{Q})$
vanishing on the $sl(N_c,{\bf C})$ variation of $\widetilde{Q}$,
as seen from the F-flatness (\ref{hk}).
Thus, the part of the Higgs branch
where the rank of $\widetilde{Q}$ is $N_c$ can be identified
with the cotangent bundle
of the (non-zero) determinant bundle of $Gr(N_c,N_f)$.
This is an open dense subset of the Baryonic branch.
Note, however, that there are vacua such that
${\rm rank}\widetilde{Q}<N_c$ and ${\rm rank}Q=N_c$ \cite{aps},
and hence the above subset is a proper subset of the moduli
space corresponding to complete Higgsing.

There is an isomorphism (as complex manifolds) between 
the determinant of $Gr(N_c,N_f)$ and that of $Gr(N_f-N_c,N_f)$.
The isomorphism can be constructed as follows.
Let $Gr(N_c,N_f)$ be realized
as the space of $N_c$ planes in a vector space $V$
of dimension $N_f$, and let $Gr(N_f-N_c,N_f)$ be realized as the space of
$N_f-N_c$ planes in its dual $V^*$.
We fix an element $v^1\wedge\cdots\wedge v^{N_f}$
of the top exterior power $\wedge^{N_f}V^*$. To an element
$w_1\wedge\cdots\wedge w_{N_c}$ in the determinant line
over the $N_c$ plane $W\subset V$ spanned by $w_1,\ldots,w_{N_c}$,
we associate an element
$i_{w_1}\cdots i_{w_{N_c}}(v^1\wedge\cdots\wedge v^{N_c})$
in the determinant line over the $N_f-N_c$ plane $W^{\perp}\subset V^*$
orthogonal to $W$. Here, $i_v$ is the interior product mapping
$q$-th exterior power of $V^*$ to $q-1$-th.
Thus, open dense subsets of the Baryonic branches
of the $SU(N_c)$ and the $SU(N_f-N_c)$ QCDs
with $N_f$ flavors are holomorphically identical.

The above discussion suggests that $N=2$ duality is only
a duality of the Baryonic branches.
It is also clear that since the D-brane moduli space that we consider 
describes only part of the Higgs branch of the $SU(N_c)$
gauge theory, we are unable in this model
to make any predictions about the the behavior of the Coulomb branch
of the $N=2$ theory
under T-duality.

The complex structure of the 
D-brane moduli space depends on the complex structure of the $K3$ surface.
On the the other hand the complex structure of the Baryonic branch of
the $N=2$ theory
on ${\bf R}^4$ 
is fixed by the D-term and F-term equations that determine the branch as
a hyperk\"ahler quotient. This seems puzzling, since we wish to identify the
Baryonic branch with the D-brane moduli space.
To this puzzle, two resolutions are possible.
One possibility is that the supersymmetric
Lagrangian field theory as we formulate it corresponds to picking one
complex structure of 
the D-brane
moduli space but there are other field theories that correspond
to picking other  complex 
structures. 
The other possibility
is that if we appropriately take the field theory limit 
the dependence on the complex structure of $K3$ disappears, and
all will yield the same result.

\subsection{Comments on $N=1$ Duality}

If the surface $S$ is rational, the gauge theory on ${\bf R}^4$ is
$N=1$ supersymmetric \cite{b}. 
By rational surface we mean a complex surface birationally
equivalent to ${\bf P}^2$.
A rational surface $S$  satisfies  $h_{1,0}(S) = h_{2,0}(S) =0$.
Consider for example the Hirzebruch  surfaces $F_n$.

As in the $K3$ case, we consider
$F_n$ as the moduli space of a 0-brane
on $F_n$ with charge vector $(0,0,-1)$. The dual to $F_n$ is
 the moduli space of vector bundles 
  with Mukai vector $v=(1,0,0)$.  
As we discussed in section 2, since the canonical class
of $S$ is non trivial, the definition of
Mukai vector (\ref{ve}) has to be modified 
in order to take into account the fact that the fermions and scalars
on the surface $S$ are
twisted. This implies that   
 bundles 
  with Mukai vector $v=(1,0,0)$ have rank one, $c_1=0$, $c_2=1$ \footnote{
  We use for the twisted case the Mukai vector 
  $v(E) = {\rm ch}(E\otimes K^{-\frac{1}{2}})\sqrt{\widehat{\rm A}(X)}$.}.
The moduli space of bundles on $F_n$
with such a  Mukai vector is the Hilbert scheme of one point
and is isomorphic (as a complex manifold) to $F_n$.
We can now follow the same steps as in the $K3$ case in order
to construct T-duality. This, however, does not lead to the 
 required exchange of 0-brane and 4-brane charges.
For the required exchange of charges, it seems that
we have to define the dual $F_n$ as
the moduli space of flat line bundles on $F_n$.
This cannot be the case since the latter
moduli space is trivial.
Similar analysis can be carried for other rational surfaces such
as blow-up of $P^2$ at points. 
As in the $F_n$ case, the results indicate that some modification
of the scenario
is needed in order to make the $N=1$ duality to work.

The duality between heterotic string theory on
$T^4$ and type IIB string theory on $K3$
was useful in order to gain an understanding of T-duality on $K3$ using
our knowledge of T-duality on $T^4$.
Similarly, it is likely that the duality between
heterotic string theory on $K3$ and 
type IIB string theory on $F_n$
(in the appropriate F-theory context) \cite{mv}
 can be used to gain an understanding of the generalization of T-duality
 on $K3$ surfaces, as constructed in this paper, to the required
T-duality on $F_n$.

\section*{Acknowledgements}
We would like to thank I.~Antoniadis, M.~Bershadsky, J.~de Boer, H.~Ooguri, B.~Pioline
and Z.~Yin for useful
discussions.
This work is supported in part by 
NSF grant PHY-951497 and DOE grant DE-AC03-76SF00098.


\newpage
 
\begin{thebibliography}{99}

\small
\parskip=0pt plus 2pt

\bibitem{b} M.~Bershadsky, A.~Johansen, T.~Pantev, V.~Sadov and C.~Vafa,
``F-Theory, Geometric Engineering and $N=1$ Dualities,'' hep-th 9612052.
\bibitem{vz}  C.~Vafa and B.~Zweibach, 
``$N=1$ Dualities of $SO$ and $USp$ Gauge Groups and T-Duality of String Theory,'' 
hep-th 9701015.
\bibitem{david} S.~Elitzur, A.~Giveon and D.~Kutasov, ``Branes and $N=1$ Duality in
String Theory,'' hep-th 9702014.
\bibitem{ov} H.~Ooguri and C.~Vafa, ``Geometry of $N=1$ Dualities in Four Dimensions,''
hep-th 9702180.
\bibitem{s} N.~Seiberg, ``Electric-Magnetic Duality in
Supersymmetric Non-Abelian 
Gauge Theories,'' hep-th 9411149.
\bibitem{k} D.~Kutasov, ``A Comment on $N=1$
 Duality in Supersymmetric Non-Abelian 
Gauge Theories,'' hep-th 9503083;\\
D.~Kutasov and A.~Schwimmer,
``On Duality in Supersymmetric Yang-Mills Theory,'' hep-th 9505004;\\
D.~Kutasov and A.~Schwimmer and N.~Seiberg, ``Chiral Rings,
Singularity Theory and Electric-Magnetic Duality,
'' hep-th 9510222.
\bibitem{HM} J.~H.~Harvey and G.~Moore, ``On The Algebras of BPS States,''
hep-th 9609017.
\bibitem{I} M.~Green, J.~H.~Harvey and G.~Moore,
``I-Brane Inflow and Anomalous Couplings
on D-Branes,'' hep-th 9605033.
\bibitem{Mukai1} S.~Mukai,
``Symplectic Structure of the Moduli Space of Sheaves on an Abelian
or $K3$ Surface,'' \invm77,84,101;
``On the Moduli Space of Bundles on $K3$ Surfaces I,''
in {\sl Vector bundles on Algebraic varieties} Tata Inst,of Fund.Research.
\bibitem{Mukai} S.~Mukai,
``Duality between $D(X)$  and $D(\widehat{X})$ with its Application
to Picard Sheaves,'' \ng81,81,153. 
\bibitem{W} E.~Witten, ``Small Instantons in String Theory,'' hep-th 9511030.
\bibitem{D} M.~R.~Douglas,
``Gauge Fields and D-branes,'' hep-th 9604198.
\bibitem{V}  C.~Vafa, ``Instantons on D-branes,'' hep-th 9512078.
\bibitem{JLi} J.~Li, ``Algebraic Geometric Interpretation of Donaldson's
Polynomial Invariants of Algebraic Surfaces'',
J. Diff. Geom. {\bf 37} (1993) 416.
\bibitem{BSV} M.~Bershadsky, V.~Sadov and C.~Vafa,
``D-Branes and Topological
Field Theories,'' hep-th 9511222.
\bibitem{KV} S.~Katz and C.~Vafa, ``Geometric Engineering of $N=1$
Quantum Field Theories,'' hep-th 9611090.
\bibitem{BB} P.~J.~Braam and P.~Van Baal,
``Nahm's Transformation for Instantons,''
\cmp122,89,267. 
\bibitem{S} H,~Schenk, ``On a Generalized Fourier  Transform of
 Instantons Over Flat Tori,''
\cmp116,88,177.
\bibitem{DK} S.~K.~Donaldson and P.~B.~Kronheimer,
``The Geometry of Four-Manifolds,''
Clarendon Press, Oxford, 1990. 
\bibitem{it} C.~Bartocci, U.~Bruzzo and D.~Hernandez Ruiperez,
``A Fourier-Mukai Transform for Stable Bundles on $K3$ Surfaces,''
alg-geom 9405006.
\bibitem{Ko} M.~Kontsevich, ``Homological Algebra of Mirror Symmetry,''
alg-geom 9411018.
\bibitem{SYZ} A.~Strominger, S.~T.~Yau and E.~Zaslow,
``Mirror Symmetry is T-Duality,''
hep-th 9606040.
\bibitem{Morrison} D.~Morrison,
``The Geometry Underlying Mirror Symmetry,''
alg-geom 9608006.
\bibitem{bb} K.~Becker, M.~Becker, D.~Morrison, 
H.~Ooguri, Y.~Oz and Z.~Yin, "Supersymmetric Cycles in Exceptional Holonomy
Manifolds and Calabi-Yau 4-Folds," 
hep-th 9608116.
\bibitem{vafa} C.~Vafa, ``Instantons on D-Branes,'' hep-th 9512078.
\bibitem{asp} P.~S.~Aspinwall,
``$K3$ Surfaces and String Duality,'' hep-th 9611137
\bibitem{aps} P.~C.~Argyres, M.~R.~Plesser and N.~Seiberg,
``The Moduli Space of Vacua of N=2 Susy QCD and Duality 
in N=1 Susy QCD,''  hep-th 9603042.
\bibitem{Anto} I.~Antoniadis and B.~Pioline,
``Higgs Branch, HyperK\"ahler quotient and duality in
SUSY N=2 Yang-Mills theories,''
hep-th 9607058.
\bibitem{mv} D.~Morrison and C.~Vafa, ``Compactification of F-Theory on Calabi-Yau 
Threefolds -I,II,''
hep-th 9602114, 9603161.





\end{thebibliography}
\end{document}